\documentclass[prb,aps,twocolumn,showpacs,preprintnumbers,superscriptaddress]{revtex4-1}
\usepackage{amsmath}
\usepackage{amsfonts}
\usepackage{amssymb}
\usepackage{graphicx}
\usepackage{bm}

\pdfoutput=1

\def\bands{\sigma_1,\sigma_2} 

\begin{document}

\title{Screening in gated bilayer graphene via variational calculus}

\author{M. M. Fogler}
\affiliation{Department of Physics, University of
California San Diego, 9500 Gilman Drive, La Jolla, CA 92093}

\author{E. McCann}
\affiliation{Department of Physics, Lancaster University, Lancaster, LA1
4YB, UK}

\begin{abstract}

We analyze the response of bilayer graphene to an external transverse electric field using a variational method. A previous attempt to do so in a recent paper by Falkovsky [Phys. Rev. B \textbf{80}, 113413 (2009)] is shown to be flawed. Our calculation reaffirms the original results obtained by one of us [E.~McCann, Phys. Rev. B \textbf{74}, 161403(R) (2006)] by a different method. Finally, we generalize these original results to describe a dual-gated bilayer graphene device.

\end{abstract}


\maketitle


The physics of monolayer and bilayer graphene has been a subject of much interest recently.~\cite{Castroneto2009tep} A unique feature of bilayer graphene (BLG) is its tunable band structure: its bandgap depends on the external electric field, which can be controlled by doping or gating. This effect was first analyzed theoretically~\cite{McCann2006lld, Castro2007bbg, McCann2006agi} and recently studied experimentally.~\cite{Ohta2006ces, Castro2007bbg, Oostinga2007gii, Henriksen2008cri, Li2008dcd, Zhang2009doo, Mak2009ooa, Kuzmenko2009iso, Kuzmenko2009dot, Kuzmenko2009gti, Zhao2010sbi,  Kim2009qhe}

\begin{figure}
\includegraphics[width=0.30\textwidth]{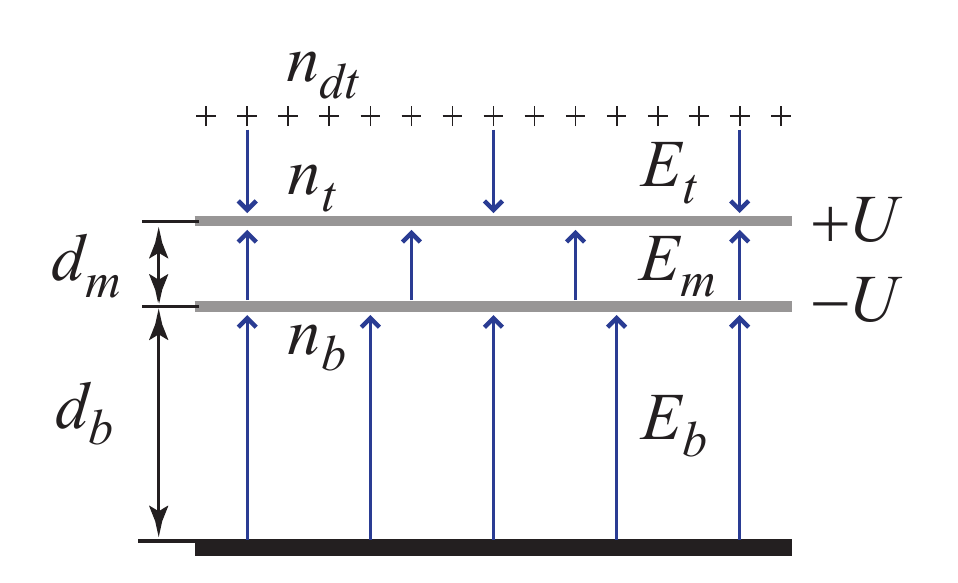}
\caption{(Color online) Device schematic. The BLG is shown as two horizontal lines in the middle. The thicker line at the bottom represents the gate. The arrows show the electric field in the system and the crosses depict positive charge density $n_{dt}$ on (or on and above) the top layer of the BLG.
\label{fig:device}
}
\end{figure}

The subject of this paper is the value of the gap at the Brillouin zone corners, which we denote by $2|U|$. As pointed out previously,~\cite{McCann2006lld, Castro2007bbg} $2 U$ coincides with the electrostatic energy difference per electron in the two layers,
\begin{equation}
2 U = e E_m d_m\,,   \quad d_m = 0.33\,\text{nm}\,,
\label{eqn:2U}
\end{equation}
where $E_m$ is the component of the electric field inside the BLG directed from the bottom to the top layer and $d_m$ is the interlayer spacing. This field depends not only on the external gate voltages but also on the induced electron densities $n_t$ and $n_b$ of the layers, see Fig.~\ref{fig:device}. For a given chemical potential $\mu$, these densities, and thus the total electron density of BLG $n = n_t + n_b$ are nonlinear functions of $U$. In general, they can be calculated only numerically.~\cite{McCann2006agi, Nilsson2006eei, Min2007ait, McCann2007eib, Castro2007bbg, Nilsson2008epo, Zhang2008dot} However, in a range $|U| \ll |\mu| \ll \gamma_1$ an asymptotic analytical formula for the interlayer bias was derived by one of us~\cite{McCann2006agi}
\begin{equation}
2 U \simeq \gamma_1 \frac{N - 2 N_{dt}}
       {\Lambda_\text{Mc}^{-1} + |N| - \frac{1}{2} \ln |N|}\,,
  \quad N \equiv \frac{n}{n_*}\,.
\label{eqn:U_McCann}
\end{equation}
The derivation was done within the commonly used approximation that neglects certain small electronic structure parameters $\gamma_3$, $\gamma_4$, and $\Delta^\prime$ (for their physical meaning and a discussion of their numerical values, see Ref.~\onlinecite{Zhang2008dot}.) The parameters that are retained include the interlayer hopping energy $\gamma_1 = 0.41\,\text{eV}$ and the Fermi velocity of \emph{monolayer} graphene $v = 1.0 \times 10^8\,\text{cm}/\text{s}$, which define the characteristic density scale in the problem
\begin{equation}
n_* = \frac{\gamma_1^2}{\pi \hbar^2 v^2}
     = 1.2 \times 10^{13} \,\text{cm}^{-2}\,.
\label{eqn:n_*}
\end{equation}
The remaining notations in Eq.~\eqref{eqn:U_McCann} are
\begin{equation}
N_{dt} = n_{dt} / n_*\,,
\label{eqn:N_dt}
\end{equation}
which is the scaled background density of positive charge $n_{dt}$ on the top layer (or above it, see below) and
\begin{equation}
\Lambda_\text{Mc} = \frac{2 \pi e^2 d_m n_*}{\kappa_m \gamma_1} \sim 1\,,
\label{eqn:Lambda_M}
\end{equation}
which is the dimensionless strength of the interlayer screening, with $\kappa_m \approx 1$ being the effective dielectric constant of the medium between the layers. The screening is particularly significant in the narrow gap regime. If $N_{dt} = 0$, it is realized in the limit $n \to 0$. According to Eq.~\eqref{eqn:U_McCann}, the screened field $E_m$ is suppressed compared to the external field $E_b$ by a divergent logarithmic factor [see Fig.~\ref{fig:device} and also Eq.~\eqref{eqn:2U_equation_v2} below]. As a result, $2U$ has a superlinear dependence on $n$.

Formula~\eqref{eqn:U_McCann} was recently challenged by Falkovsky.~\cite{Falkovsky2009sig} Under the same conditions and assumptions, he obtained the simple linear law:
\begin{equation}
2 U = c \gamma_1 N\,, \quad c = \frac{2}{\sqrt{6.2784^2 - 1}} = 0.3227\,.
\label{eqn:U_Falkovsky}
\end{equation}
The goal of this paper is to show that Ref.~\onlinecite{Falkovsky2009sig} contains egregious mistakes. Once they are corrected, Eq.~\eqref{eqn:U_McCann} is recovered.

Some of these mistakes are apparent on physical grounds. One key formula, Eq.~(8) of Ref.~\onlinecite{Falkovsky2009sig}, states (in our notations) that the BLG polarization
\begin{equation}
\Delta n \equiv n_t - n_b
\label{eqn:Deltan_def}
\end{equation}
is positive for $U, \mu > 0$, i.e., the BLG is polarized against, not along the electric field. In principle, such a phenomenon can arise if a system has a negative compressibility~\cite{Bello1981dol, Dolgov1981oaa, Eisenstein1994cot} due to exchange and correlations (see the discussion at the end). However, neither exchange nor correlation effects are included in the theory of Ref.~\onlinecite{Falkovsky2009sig}.

How the author of Ref.~\onlinecite{Falkovsky2009sig} is able to eventually arrive at Eq.~\eqref{eqn:U_Falkovsky} is also unclear to us. Even if we take for granted the basic Eqs.~(8)--(10) of Ref.~\onlinecite{Falkovsky2009sig} and dutifully follow the suggested steps of the derivation, we are still unable to reproduce the later equations. For example, on the left-hand side of Eq.~(13) of Ref.~\onlinecite{Falkovsky2009sig} we get the \emph{top} layer density instead of the \emph{bottom} one stated there, and on the right-hand side we get an extra minus sign. With none of the choices for the sign or the layer can we get from Eq.~(13) to Eq.~(14) of Ref.~\onlinecite{Falkovsky2009sig}.

To settle the matter decisively we carry out below our own derivation using the method proposed by Falkovsky in Ref.~\onlinecite{Falkovsky2009sig}. Its idea is to treat $U$ and $\mu$ as variational parameters and get their actual values by minimizing the total energy density $E$ of the system at a fixed $n$. Note that $\mu$ and $U$ are admissible variational parameters if the Jacobian of the transformation from $n_t$ and $n_b$ to $\mu$ and $U$ is non-degenerate:
\begin{equation}
 \frac{\partial(n_t, n_b)}{\partial (\mu, U)}
 \equiv \frac{\partial n_t}{\partial \mu} \frac{\partial n_b}{\partial U}
 - \frac{\partial n_t}{\partial U} \frac{\partial n_b}{\partial \mu}
 \neq 0\,.
\label{eqn:Jacobian}
\end{equation}
This condition can be verified from the final results. We show that the variational method leads back to Eq.~\eqref{eqn:U_McCann} and that all the intermediate results make physical sense.

Let us proceed. In the mean-field approximation $E$ is the sum of the kinetic and the Hartree interaction terms:
\begin{equation}
E = E_\text{kin} + E_\text{H}\,,
\label{eqn:E}
\end{equation}
The Hartree term is straightforward:
\begin{equation}
E_\text{H} = \frac12 e^2 c_b (n - n_{db}^0)^2
 + \frac{2 \pi e^2 d_m}{\kappa_m} (n_t - n_{dt})^2\,,
\label{eqn:E_H}
\end{equation}
which can be understood as the energy of two parallel-plate capacitors in series. Here $c_b = 4 \pi \kappa_b / d_b$ is the capacitance per unit area between the BLG and the bottom gate, $\kappa_b$ is the dielectric constant below the BLG, and $n_{db}^0$ is the density of additional positive background charge charge on the bottom layer (not shown in Fig.~\ref{fig:device}). For $n_{db}^0 = 0$, Eq.~\eqref{eqn:E_H} agrees with Eq.~(10) in Ref.~\onlinecite{Falkovsky2009sig}.

The kinetic term $E_\text{kin}$ requires a little care. The author of Ref.~\onlinecite{Falkovsky2009sig} seems to assume that it coincides with the sum of the occupied single-particle energies
\begin{equation}
E_s = \frac{1}{A} \sum\limits_\alpha \Theta(\mu - \epsilon_\alpha) \epsilon_\alpha
\equiv \sum \epsilon_\alpha\,,
\label{eqn:E_s}
\end{equation}
where $\alpha$ is a short-hand notation for all quantum numbers, $\Theta(x)$ is the unit step-function, and $A$ is the area of the system. In fact, each energy $\epsilon_\alpha$, which is an eigenvalue of the BLG Hamiltonian
$H = H_\text{kin} + \Phi$ also contains a potential term:
\begin{equation}
  \epsilon_\alpha = \langle\alpha | H_\text{kin} | \alpha \rangle + \langle\alpha | \Phi | \alpha \rangle\,.
\label{eqn:H}
\end{equation}
Here $\Phi$ stands for the electrostatic energy that is equal to $\pm U$ in the top (bottom) layer. Lumping together $E_s$ and $E_\text{H}$ is incorrect as it leads to double-counting of the interaction energy. Instead, the proper formula is
\begin{equation}
E_\text{kin} = \sum
               \langle\alpha | H_\text{kin} | \alpha \rangle
             = E_s - U \Delta n\,.
\label{eqn:E_kin}
\end{equation}
Next, using the Hellmann-Feynman theorem, we get
\begin{equation}
\Delta n = \sum
             \left\langle\alpha \left| \frac{\partial H}{\partial U} \right| \alpha \right\rangle
           = \frac{\partial}{\partial U} (E_s - \mu n)\,.
\label{eqn:dn}
\end{equation}
Note also the useful relations
\begin{equation}
n = n_t + n_b = \sum\, 1\,,
\quad
\frac{\partial}{\partial \mu} E_s = \mu \frac{\partial n}{\partial \mu}\,.
\label{eqn:n}
\end{equation}
Therefore,
\begin{equation}
n_{t, b} = \frac{n \pm \Delta n}{2} = \frac12 \left[
n \pm \frac{\partial}{\partial U} (E_s - \mu n)
\right]\,.
\label{eqn:n_tb}
\end{equation}
The correct variational functional to minimize is
\begin{equation}
\Omega = E_\text{H} + E_s - U \frac{\partial}{\partial U} (E_s - \mu n)
       - \xi n\,,
\label{eqn:Omega}
\end{equation}
where $\xi$ is a Lagrange multiplier. The sought minimum is determined by the equations $\partial \Omega / \partial \mu = \partial \Omega / \partial U = 0$. Using Eqs.~\eqref{eqn:n_tb} and \eqref{eqn:n}, one can bring them to the form
\begin{align}
&C \frac{\partial n_t}{\partial \mu} + D \frac{\partial n}{\partial \mu} = 0\,,
\quad
C \frac{\partial n_t}{\partial U} + D \frac{\partial n}{\partial U} = 0\,,
\label{eqn:CD}\\
&C \equiv \frac{\partial}{\partial n_t} E_\text{H}(n_t, n) - 2 U\,,
\label{eqn:C}\\
&D \equiv \mu + U + \frac{\partial}{\partial n} E_\text{H}(n_t, n) - \xi\,.
\label{eqn:D}
\end{align}
In view of Eq.~\eqref{eqn:Jacobian}, the system of linear equations for $C$ and $D$ has only the trivial solution $C = D = 0$, and so
\begin{equation}
2U = 4 \pi e^2 ({d_m} / {\kappa_m}) (n_t - n_{dt})\,.
\label{eqn:U_equation}
\end{equation}
Of course, this equation can be written at the outset because it follows from Gauss's law. Indeed, it is the starting equation [Eq.~(2)] of Ref.~\onlinecite{McCann2006agi}. The closest to Gauss's law in Ref.~\onlinecite{Falkovsky2009sig} is Eq.~(13) of that paper but it has the problems described above.

Let us now compute $n_t$ by the present method to demonstrate where Ref.~\onlinecite{Falkovsky2009sig} contains another mistake. The total polarization $\Delta n$ is the sum over all four bands of BLG
\begin{equation}
\Delta n = \sum\limits_{\bands} \Delta n_{\bands}\,.
\label{eqn:Deltan}
\end{equation}
We use the signature $\bands$ to label the bands as follows: $\sigma_1 = \pm$ distinguishes the conduction and valence bands and $\sigma_2 = \pm$ labels the outer (inner) bands. The band dispersions are given by~\cite{McCann2006lld}
\begin{align}
  \epsilon_{\bands}(U, k) &= \sigma_1
  \sqrt{\frac{\gamma_1^2}{2} + \varepsilon^2(k) + U^2
    +\sigma_{2} R_\epsilon}\,,
\label{eqn:BLG_E_k} \\
  R_\epsilon &\equiv \sqrt{\frac{\gamma_1^4}{4} + \varepsilon^2(k) (\gamma_1^2 + 4 U^2) }\,,
\label{eqn:R_E}
\end{align}
where $k$ is the deviation of the quasimomentum from the nearest Brillouin zone corner and $\varepsilon(k) = \hbar v k$. Each band has $g = 4$ fold degeneracy due to spin and valley.

For simplicity, we consider the case where the Fermi surface is singly-connected and includes only the states of the $\bands = +,-$ band, so that
$n = g (\pi k_F^2) / (2 \pi)^2 = k_F^2 / \pi$ where $k_F$ is the Fermi momentum and
\begin{equation}
\mu = \epsilon_{+,-}(k_F) \simeq
\sqrt{U^2 + \frac{\varepsilon^4(k_F)}{\gamma_1^2}} \simeq
\sqrt{U^2 + \gamma_1^2 N^2}\,.
\label{eqn:mu}
\end{equation}
As explained above, if $U$ is positive, then $\Delta n$ should be negative. In
Ref.~\onlinecite{Falkovsky2009sig} however only the conduction band term,~\cite{McCann2006agi, Falkovsky2009sig}
\begin{equation}
\Delta n_{+,-} = \frac{n_* U}{\gamma_1} \ln\left(
\frac{\mu}{U} + \sqrt{1 + \frac{\mu^2}{U^2}}\right)\,,
\label{eqn:Deltan_c}
\end{equation}
is included while the much larger negative contribution $\Delta n_- \equiv \Delta n_{-,+} + \Delta n_{-,-}$ of the completely filled valence bands is neglected. This important term is given by [cf.~Eq.~\eqref{eqn:dn}]
\begin{equation}
\Delta n_- = \frac{\partial}{\partial U}\int \frac{g d^2 k}{(2\pi)^2}
\left.[\epsilon_{-,+}(k) + \epsilon_{-,-}(k)]\right|^{U > 0}_{U = 0}\,.
\label{eqn:Deltan_v}
\end{equation}
The above integral can be done exactly but we need only its leading asymptotic form:
\begin{equation}
\Delta n_{-} \simeq -\frac{n_* U}{\gamma_1}\, \ln \frac{2 \gamma_1}{U}\,,
\label{eqn:eqn:Deltan_v_as}
\end{equation}
which is in agreement with Ref.~\onlinecite{McCann2006agi}. Combined with Eqs.~\eqref{eqn:mu} and \eqref{eqn:Deltan_c}, it yields the result of the
correct sign,
\begin{equation}
\Delta n \simeq -\frac{n_* U}{\gamma_1}\, \ln \frac{2}{N + \sqrt{N^2 + (U/\gamma_1)^2}} \,.
\label{eqn:eqn:Deltan_as}
\end{equation}
Substituting it into $n_t = (n + \Delta n) / 2$ and using Eqs.~\eqref{eqn:N_dt}, \eqref{eqn:Lambda_M}, and \eqref{eqn:U_equation}, we obtain the equation for the function $U(N)$:
\begin{equation}
2 U(N) \simeq \gamma_1 \frac{N - 2 N_{dt}}{\displaystyle \Lambda_\text{Mc}^{-1} +  \frac12 \ln \frac{2}{N + \sqrt{N^2 + (U / \gamma_1)^2}}}\,.
\label{eqn:2U_equation}
\end{equation}
This formula is actually valid for arbitrary signs of $U$ and $N$. In the region
\begin{equation}
{2U(0)} / {\gamma_1} \ll N \ll 1
\label{eqn:U_McCann_region}
\end{equation}
it coincides with Eq.~\eqref{eqn:U_McCann} within the accuracy of this calculation. Here
\begin{equation}
2U(0) \simeq -\frac{2\gamma_1 N_{dt}}{\displaystyle \Lambda_\text{Mc}^{-1} +  \frac12 \ln \frac{1}{|N_{dt}|}}
\label{eqn:2U_0}
\end{equation}
is the gap at $N = 0$. Note that the gap vanishes only at one density, $N = 2 N_{dt}$, i.e., $n = 2 n_{dt}$, at which
\begin{equation}
n_t = n_b = n_{dt}\,.
\label{eqn:zero_gap}
\end{equation}
The gap remains nonzero at all other $n$, in disagreement with the claim made in Ref.~\onlinecite{Falkovsky2009sig} but in agreement with the numerical results of Ref.~\onlinecite{McCann2006agi}.

As mentioned in the beginning of this paper, the background positive charge density $n_{dt}$ does not have to reside directly on the top layer. Equation~\eqref{eqn:2U_equation} remains unchanged if some of this charge is located above the BLG, as sketched in Fig.~\ref{fig:device}. For example, for a dual-gated BLG
\begin{equation}
n_{dt} = n_{dt}^0 + \frac{c_t}{e^2} (e V_t - \mu + U)\,,
\label{eqn:n_dt}
\end{equation}
where the first term is the fixed donor density on the top layer and the second term is the (tunable) charge density on the top gate. Parameter $c_t$ is the capacitance between the top gate and the BLG per unit area. Similarly, the total density is given by
\begin{equation}
n =  n_{dt}^0 + n_{db}^0 + \frac{c_t - c_b}{e^2}U + \sum\limits_{a = t, b} \frac{c_a}{e^2}(eV_a - \mu)\,,
\label{eqn:n_from_V}
\end{equation}
where $V_b$ is the voltage on the bottom gate. The chemical potential $\mu$ enters these equations because the measured ``voltages'' $V_b$ and $V_t$ of the two gates are not simply the electrostatic but instead the electrochemical potential differences between these gates and the BLG, cf.~Refs.~\onlinecite{Bello1981dol} and \onlinecite{Eisenstein1994cot}. However for distant gates,
\begin{equation}
d_a \gg d_m \frac{\kappa_a}{2 \Lambda_\text{Mc} \kappa_m}\,,
\label{eqn:distant_gates}
\end{equation}
one can in the first approximation neglect $U$ and $\mu$ compared to $e V_b$ and $e V_t$, leading to the simplified equations
\begin{align}
2 U &= \frac{e E_0 d_m}{\displaystyle
 1 +  \frac{\Lambda_\text{Mc}}{2} \ln \frac{2}{N + \sqrt{N^2 + (U / \gamma_1)^2}}}\,,
\label{eqn:2U_equation_v2}\\
E_0 &\simeq \frac{2\pi e}{\kappa_m} \left(n_{db}^0 - n_{dt}^0
+ \frac{c_b V_b}{e} - \frac{c_t V_t}{e}\right),
\label{eqn:E_0}\\
N &=\frac{n}{n_*} \simeq \frac{1}{n_*} \left(
                         n_{db}^0 + n_{db}^0
                         + \frac{c_t V_t}{e}
                         + \frac{c_b V_b}{e}\right)\,,
\label{eqn:N_from_V}
\end{align}
which may be useful in experimental practice. Note however that unavoidable disorder creates additional corrections to these expressions,~\cite{Zhang2008dot} which can be as large as $30\%$.


In closing, we comment on electron-electron exchange and correlations. These many-body effects have been predicted to generate a band gap~\cite{Nilsson2006eei, Min2007pmi, Zhang2009eei, Nandkishore2009fte} and the corresponding polarization $\Delta n \neq 0$ even in the absence of the external electric field. Although neglected in our mean-field theory, exchange and correlations can be incorporated by introducing a suitable self-energy difference between the layers, which is a function of $N$ and $U$. In fact, we already have a similar parameter in our formalism --- this is our $N_{dt}$. We intend to investigate this interesting problem in a future work.

This work is supported by the NSF Grant DMR-0706654 (MMF) and by EPSRC First Grant EP/E063519/1 (EM).

\input{Falkovsky_bbl}
\end{document}